\documentclass[a4paper,11pt]{article}
\usepackage{pos}

\newcommand{\mrm}[1]{_{\rm #1}}
\renewcommand{\d}{{\rm d}}

\title{BlackHawk: A tool for computing Black Hole evaporation}
\ShortTitle{BlackHawk: A tool for computing Black Hole evaporation}

\author*[a]{J\'er\'emy Auffinger}
\author[a,b,c]{Alexandre Arbey}

\affiliation[a]{Univ Lyon, Univ Claude Bernard Lyon 1, CNRS/IN2P3, IP2I Lyon, UMR 5822, F-69622, Villeurbanne, France}

\affiliation[b]{Institut Universitaire de France (IUF), 103 boulevard Saint-Michel, 75005 Paris, France}

\affiliation[c]{Theoretical Physics Department, CERN, CH-1211 Geneva 23, Switzerland\\~}

\emailAdd{j.auffinger@ipnl.in2p3.fr}
\emailAdd{alexandre.arbey@ens-lyon.fr}

\abstract{We present the public code \texttt{BlackHawk}, a powerful tool to compute the evaporation of any distribution of Schwarzschild and Kerr Black Holes and the emission spectra of Hawking radiation.}

\FullConference{%
  Tools for High Energy Physics and Cosmology - TOOLS2020\\
  2-6 November, 2020\\
  Institut de Physique des 2 Infinis (IP2I), Lyon, France}

\begin{document}

\maketitle

\section{Introduction}

The nature of dark matter (DM) is a long standing problem in modern cosmology. For now, particle physics candidates explaining the missing mass have not been observed, neither in direct (see e.g. \cite{Tao2020} for an overview) nor indirect searches (see e.g. \cite{Perez2020} for a recent review). On the other hand, the concept of primordial Black Holes (PBHs) which originates from the 60s \cite{Zeldovich1967} has raised revived interest in the last decade. The recent detection of gravitational waves from the mergers of binary BHs by the LIGO/Virgo Collaborations \cite{LIGO2019} and the observation of the shadow of the supermassive BH at the center of M87 \cite{EHT2019} have shown that BHs are ubiquitous in our universe. The observed BHs are either of stellar mass or supermassive. There is a never-ending discussion about their origin: most, or at least some of them may be primordial. Many astrophysical constraints limit the fraction of DM that PBHs could represent, see for example \cite{Carr2020} and references therein for a long review, and \cite{Green2020} for a more concise one. Recent work focuses on the nature of LIGO/Virgo BHs and the discrimination between the primordial and stellar origin using mass distribution or spin measurements \cite{Khalouei2020,Boehm2020,Sobrinho2020,Wong2020,Garcia2020,Burke2020,DeLuca2020}. Most of those studies consider PBHs with very low spin formed during the radiation domination era, but it has been shown that PBHs could have a very high spin if formed during a transient matter dominated era, evading Thorne's limit \cite{Arbey2020_spin}.

The most striking characteristics of BHs is that they emit radiation close to the blackbody one and thus slowly evaporate away by losing mass, angular momentum and -- putative -- electric charge. This phenomenon first described by Hawking \cite{Hawking1975} led to the idea that BHs could be detected via the radiation they emit when evaporating \cite{Hawking1974}. Hawking radiation is sizeable only for small BHs of less than $\sim 10^{16}-10^{19}\,$g, leading to radiation in the MeV$\,-\,$keV band. This points directly towards PBHs since stellar mass BHs are heavier than $\sim 2\,M_\odot$ due to the Tolman-Oppenheimer-Volkoff limit.

PBHs may have formed in the early universe through the collapse of strong density inhomogeneities $\Delta\rho/\rho\gtrsim 1$ resulting from quantum fluctuations during inflation, from the collapses of topological defects (cosmic strings, domain walls), or from bubble collisions during a first-order phase transition (see \cite{Carr2020} for an extended discussion). The mass of PBHs formed in the early universe is linked to the content of the Hubble radius at the time of formation
\begin{equation}
    M^{\rm init}\mrm{PBH}(t) \sim 10^{15}\left( \dfrac{t}{10^{-23}\,{\rm s}} \right)\,{\rm g}\,,
\end{equation}
which corresponds approximately to the Hubble mass at time $t$. Evaporation equations show that (non-rotating) BHs have a lifetime linked to their mass by
\begin{equation}
    \tau(M^{\rm init}\mrm{PBH}) \sim 10^{64}\left( \dfrac{M^{\rm init}\mrm{PBH}}{M_\odot} \right)^3\,{\rm yr}\,. \label{eq:lifetime}
\end{equation}
Thus PBHs with a mass $M\mrm{PBH}^{\rm init}\lesssim 10^{15}\,$g have completely evaporated by now. These PBHs may have however left imprints of their evaporation in cosmological observations. Anyway, they cannot represent a sizeable fraction of DM. Heavier PBHs, on the other hand, are almost eternal compared to the age of the universe due to the power $3$ dependence in Eq.~\eqref{eq:lifetime}. If they are of asteroid mass $M\mrm{PBH}^{\rm init} \lesssim 10^{19}\,$g, then their Hawking radiation may be detected by astrophysical observation and they could represent a fraction, if not all, of DM.

The above discussion shows that there is a need for constraints on PBH DM linked to observations in order to infirm or confirm this scenario. Alternatively, mixed models assuming the existence of more than one DM component (e.g. WIMPs and PBHs \cite{Carr2020_mixed}) can also be constrained using Hawking radiation. For now, there still exists an open window $10^{17}\,{\rm g}\lesssim M\mrm{PBH}\lesssim 10^{21}\,{\rm g}$ in the asteroid and sublunar mass range for PBHs to represent all of DM \cite{Carr2020}. PBHs are also quite unconstrained in the lowest possible mass range $10^{-1}\,{\rm g}\lesssim M\mrm{PBH}\lesssim 10^{9}\,{\rm g}$ as there is no known cosmological observable from those times, although gravitational astronomy may soon provide new possibilities \cite{Inomata2020,Hooper2020}. This is the general context in which the public tool \texttt{BlackHawk} \cite{BlackHawk} has been created.

In section \ref{sec:HR}, we present rapidly the fundamental equations of Hawking radiation, in section \ref{sec:blackhawk} we discuss in some detail the capabilities of \texttt{BlackHawk}, in section \ref{sec:constraints} we present some examples of the work that can be done with \texttt{BlackHawk} and we conclude in section \ref{sec:prospects}.

\section{Hawking radiation}
\label{sec:HR}

Hawking radiation \cite{Hawking1975} is a semi-classical phenomenon by which pairs of particles and anti-particles are created in the vicinity of the horizon of a BH and can get causally separated by this horizon. There results a net flux of energy at spatial infinity. The creation of pairs at the horizon is a purely gravitational phenomenon whose thermodynamics is directed by the effective temperature of the BH\footnote{In the following, we use natural units for which $G = \hbar = c = k\mrm{B} = 1$.}
\begin{equation}
    T \equiv \dfrac{\kappa}{2\pi}\,,
\end{equation}
where $\kappa$ is the surface gravity. Thus, the rate of Hawking radiation is simply given by the blackbody law
\begin{equation}
    \dfrac{\d^2 N_i}{\d t\d E} = \sum_{\rm dof} \dfrac{\Gamma_i(E,M,x_j)/2\pi}{e^{E^\prime(E,x_j)/T}-(-1)^{2s_i}}\,,
\end{equation}
where
\begin{itemize}
    \item $\d^2N_i/\d t\d E$ is the number of particles of type $i$ emitted per unit time $t$ and energy $E$;
    \item $\Gamma_i$ is the greybody factor describing the probability that a particle generated at the horizon escapes at infinity;
    \item $E^\prime$ is the total energy of the particles corrected for e.g. horizon rotation or electric charge;
    \item $s_i$ is the particle spin;
    \item $x_j$ are all the internal degrees of freedom of the BH, namely angular momentum $a^*$ and charge $Q$ following the no-hair theorem;
    \item the sum runs over the internal degrees of freedom of the particles (helicity, color, \emph{etc}).
\end{itemize}

In this formula, the most important component is the greybody factors which encode the discrepancy between BH radiation and pure blackbody radiation. It is evaluated by evolving a plane wave radially from the BH horizon to the spatial infinity and comparing the amplitudes of the wave at these two extreme positions
\begin{equation}
    \Gamma_i(E,M,x_j) \equiv \left| \dfrac{Z_\infty}{Z\mrm{hor}} \right|^2\,.
\end{equation}

Hawking radiation yield thus depends on the spin of the emitted particle, and on the other internal degrees of freedom such as its charge and angular momentum. Couplings between the particle and the BH electric charges, or between the particle and the BH angular momentum \cite{Page1,Page2,Page3} give rise to interesting phenomena such as superradiance \cite{Brito2015}. Although numerical and analytical estimates existed for the greybody factors for various BH solutions (\cite{MacGibbon1990} is often cited), precise numerical computations of these factors lacked in the literature.

We can deduce the time evolution of BHs by integrating over the energy and momentum loss
\begin{align}
	&f(M,a^*) \equiv -M^2 \dfrac{\d M}{\d t} = M^2\int_{0}^{+\infty} \sum_i\sum_{\rm dof.} \dfrac{E}{2\pi}\dfrac{\Gamma_i(E,M,a^*)}{e^{E^\prime/T}\pm 1} \d E\,, \label{eq:page1} \\
	&g(M,a^*) \equiv -\dfrac{M}{a^*} \dfrac{\d J}{\d t} = \dfrac{M}{a^*}\int_{0}^{+\infty} \sum_i\sum_{\rm dof.}\dfrac{m}{2\pi} \dfrac{\Gamma_i(E,M,a^*)}{e^{E^\prime/T}\pm 1}\d E\,, \label{eq:page2}
\end{align}
where $J\equiv a^* M^2$ is the angular momentum of the BH and $m$ that of the particle. These Page factors allow to write differential equations for the mass and spin of the BH
\begin{align}
	&\dfrac{\d M}{\d t} = -\dfrac{f(M,a^*)}{M^2}\,, \\
	&\dfrac{\d a^*}{\d t} = \dfrac{a^*(2f(M,a^*) - g(M,a^*))}{M^3}\,,
\end{align}
where the mass evolution equation indeed shows the behaviour of Eq.~\eqref{eq:lifetime} if $f(M,a^*)$ is taken as a constant.

\section{\texttt{BlackHawk} content}
\label{sec:blackhawk}

\texttt{BlackHawk} has been designed as a public tool to compute the Hawking radiation emitted by a variety of BH solutions, with any initial mass and spin distribution, and to provide spectra of final particles that can be compared to astrophysical observations\footnote{Many more details can be found in the \texttt{BlackHawk} manual \cite{BlackHawk}.}.

Various PBHs distributions taken from the literature have been implemented in \texttt{BlackHawk}, like the monochromatic one with a single BH mass and spin, and the log-normal one which is the result of PBH production by a narrow peak in the power spectrum of primordial density fluctuations. A lot of efforts have been put in the computation of the greybody factors for Schwarzschild (non-rotating, uncharged) and Kerr (rotating, uncharged) BHs, which are \emph{per se} interesting results. These greybody factors are used to compute the Page factors Eq.~\eqref{eq:page1} and Eq.~\eqref{eq:page2}, which are then used to compute the mass and spin evolution of the distribution of BHs.

Hawking radiation is described in terms of Standard Model (SM) particle emission. Of course, any particle beyond the SM would be emitted by the BH horizon, such as any particle DM candidate, or the putative graviton (included in \texttt{BlackHawk}). On the other hand, astrophysical observations are based on the detection of stable particles, or at least on the imprints left in some cosmological era by stable particles at that timescale. Thus, particle physics codes (\texttt{PYTHIA} \cite{Pythia2015} and \texttt{HERWIG} \cite{Herwig2020}) are used to hadronize and decay the SM particles to stable ones\footnote{Stable in the sense of cosmologically stable, or stable at the time of BBN for example.}. In \texttt{BlackHawk}, they come in the form of tabulated transfer functions from primary SM particles to secondary stable particles.

\section{Primordial Black Hole detection and constraints}
\label{sec:constraints}

\begin{figure}
    \centering
    \includegraphics[width = 0.5\textwidth]{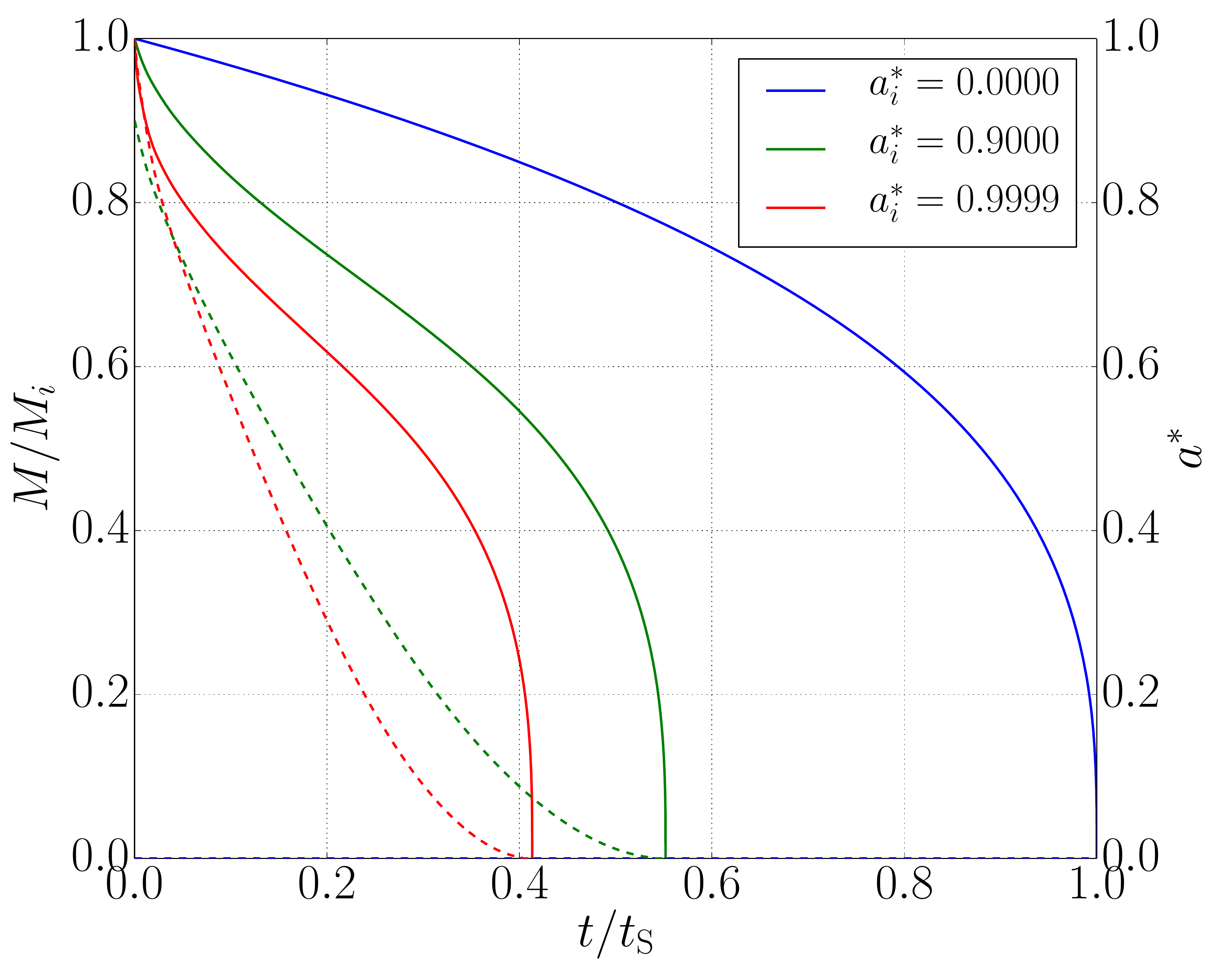}
    \caption{Evolution of a BH mass and spin following Eqs.~\eqref{eq:page1} and \eqref{eq:page2}. Taken from \cite{Arbey2020_spin}.}
    \label{fig:spin}
\end{figure}

\texttt{BlackHawk} is a versatile tool. The tabulated values of greybody factors, Page factors, the tabulated transfer functions between primary and secondary particles can be used outside the main code for e.g. BH evolution study \cite{Arbey2020_spin} (see Fig.~\ref{fig:spin}) or DM indirect searches (on the same model as for the tables of M. Cirelli \emph{et al.} \cite{Cirelli2011}). Of course, the main aim of \texttt{BlackHawk} is to provide spectra for constraining the abundance of PBHs in the universe, or for predicting signals emitted by known or putative astrophysical objects. Here we give a short overview of the work that has been done so far.

First, \texttt{BlackHawk} can predict the instantaneous spectrum of a population of PBHs in our Galaxy. If PBHs represent a substantial fraction or all of DM, they are expected to be distributed as DM in the galactic halo, following for example the Navarro-Frenk-White (NFW) profile given by \cite{Navarro1996}
\begin{equation}
    \rho\mrm{PBH}(r) = \rho\mrm{DM}(r) \sim \dfrac{1}{(r/r_s)(1+r/rs)^2}\,,
\end{equation}
where $r_s$ is a characteristic radius. Then, if PBHs are monochromatic, i.e. with the same unique mass, it is easy to deduce the flux of particles of type $i$ on an Earth or space based detector resulting from the integration over the galactic bulk
\begin{equation}
    \dfrac{\d F_i}{\d E} = \int \dfrac{\d\Omega}{4\pi}\int \d l \dfrac{\rho\mrm{PBH}(l)}{M\mrm{PBH}} \dfrac{\d^2 N_i}{\d t\d E}\,,
\end{equation}
where $l$ is the line of sight coordinate. Of course, this flux depends on the precise distribution of DM, and deviations from NFW induce modified constraints/signals. For example, \texttt{BlackHawk} can predict the density of high-energy particles emitted by PBHs at the galactic center. These particles evolve in a strong magnetic field and thus produce synchrotron radiation in the radio band, that can be used to constrain the amount of PBHs \cite{Chan2020}. PBHs of $M\mrm{PBH}\lesssim 10^{17}\,$g emit electrons and positrons. The annihilation of the $e^\pm$ in the galactic center may contribute to the $511\,$keV line measured in the direction of the galactic center \cite{Laha2019}. On the same level, the INTEGRAL measure of the soft gamma-ray spectrum of the Galaxy sets constraints on the abundance of light PBHs \cite{Laha2020,Laha2020_2} and neutrino scintillators such as JUNO are sensitive to (anti)neutrinos \cite{Laha2020_3,Wang2020}.

\begin{figure}
    \centering
    \includegraphics[width = \textwidth]{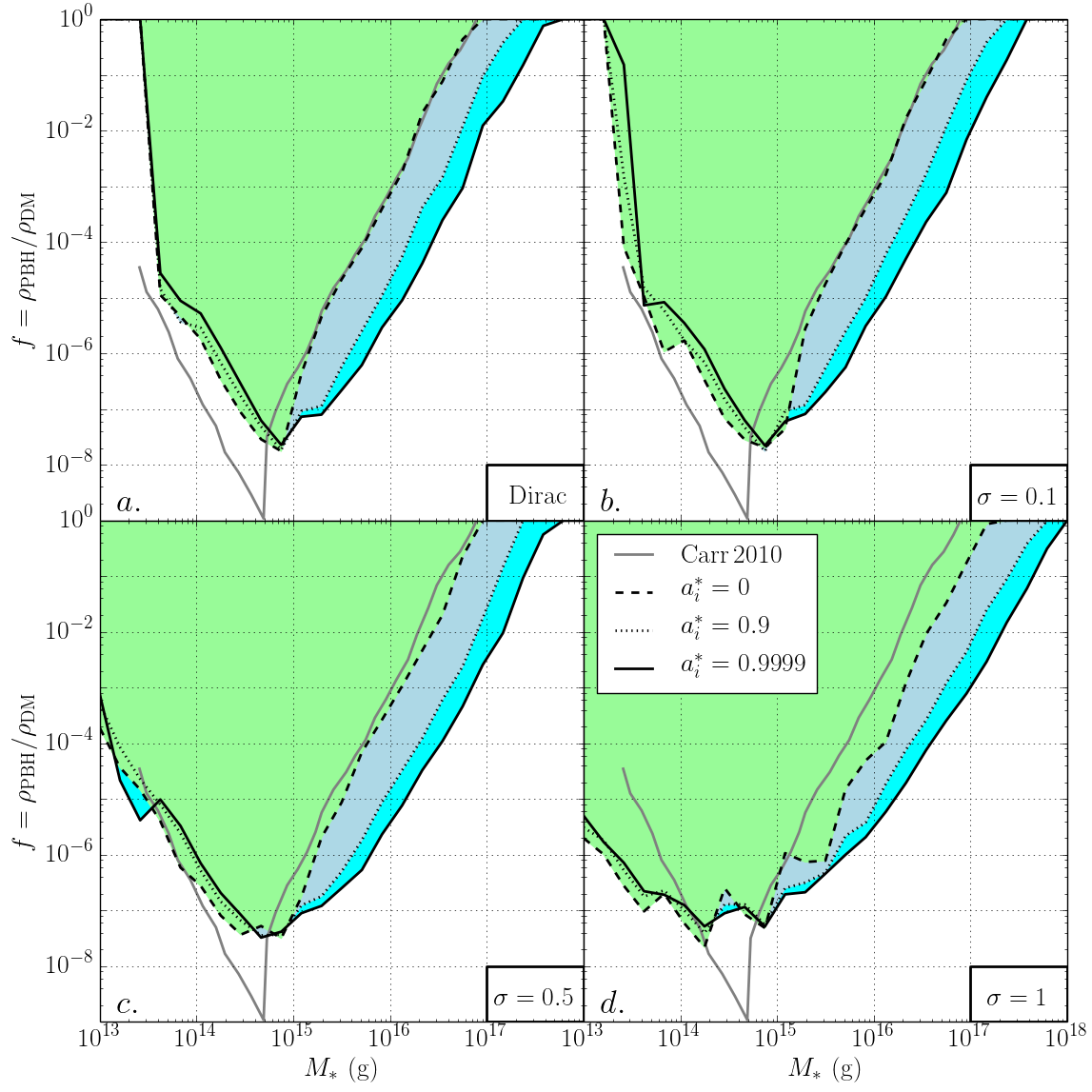}
    \caption{Constraints on the DM fraction composed of PBHs from the isotropic extragalactic gamma-ray background. Taken from \cite{Arbey2020_IGRB}.}
    \label{fig:gamma}
\end{figure}

Second, \texttt{BlackHawk} can be used to constrain the cosmological amount of PBHs. If PBHs represent a substantial fraction of DM, then they can explain both the cosmological DM and galactic DM. For that, \texttt{BlackHawk} can compute the emission of particles throughout the whole BH lifetime, and integrating this emission, taking the cosmological redshift into account, gives access to the isotropic background flux of e.g. (soft) gamma-rays \cite{Ballestros2020,Arbey2020_IGRB}
\begin{equation}
    I \equiv E \dfrac{\d F_\gamma}{\d E} = \dfrac{1}{4\pi} n\mrm{PBH}(t_0)E \int_{t\mrm{min}}^{t\mrm{max}}(1 + z) \dfrac{\d^2 N_\gamma}{\d t\d E}((1 + z)E)\d t\,,
\end{equation}
where $n\mrm{PBH}(t_0)$ is the number density of PBHs today, $z(t)$ is the redshift and $\d^2 N_\gamma/\d t\d E$ is the secondary spectrum of photons. The results of this computation, taking the PBH spin into account, are presented in Fig.~\ref{fig:gamma}. This prediction has been adapted to PBHs in scenarios with large extra spatial dimensions \cite{Johnson2020}, opening an interesting window on the ability of \texttt{BlackHawk} to probe alternative gravitational theories.

Third, \texttt{BlackHawk} can be used together with Big-Bang nucleosynthesis (BBN) codes to provide constraints on very light PBHs that evaporate when the light elements are forged at the beginning of the universe. Indeed, energetic particles emitted by $M\mrm{PBH}\sim 10^{9}-10^{13}\,$g PBHs can change the neutron to proton ratio at the onset of BBN, induce nuclear reactions during BBN or photo-dissociate the final light elements after BBN. Hence, the modified Deuterium and/or Helium abundances can provide strong constraints on the abundance of light PBHs at the BBN epoch \cite{Luo2020}.

\begin{figure}
    \centering
    \includegraphics[width = \textwidth]{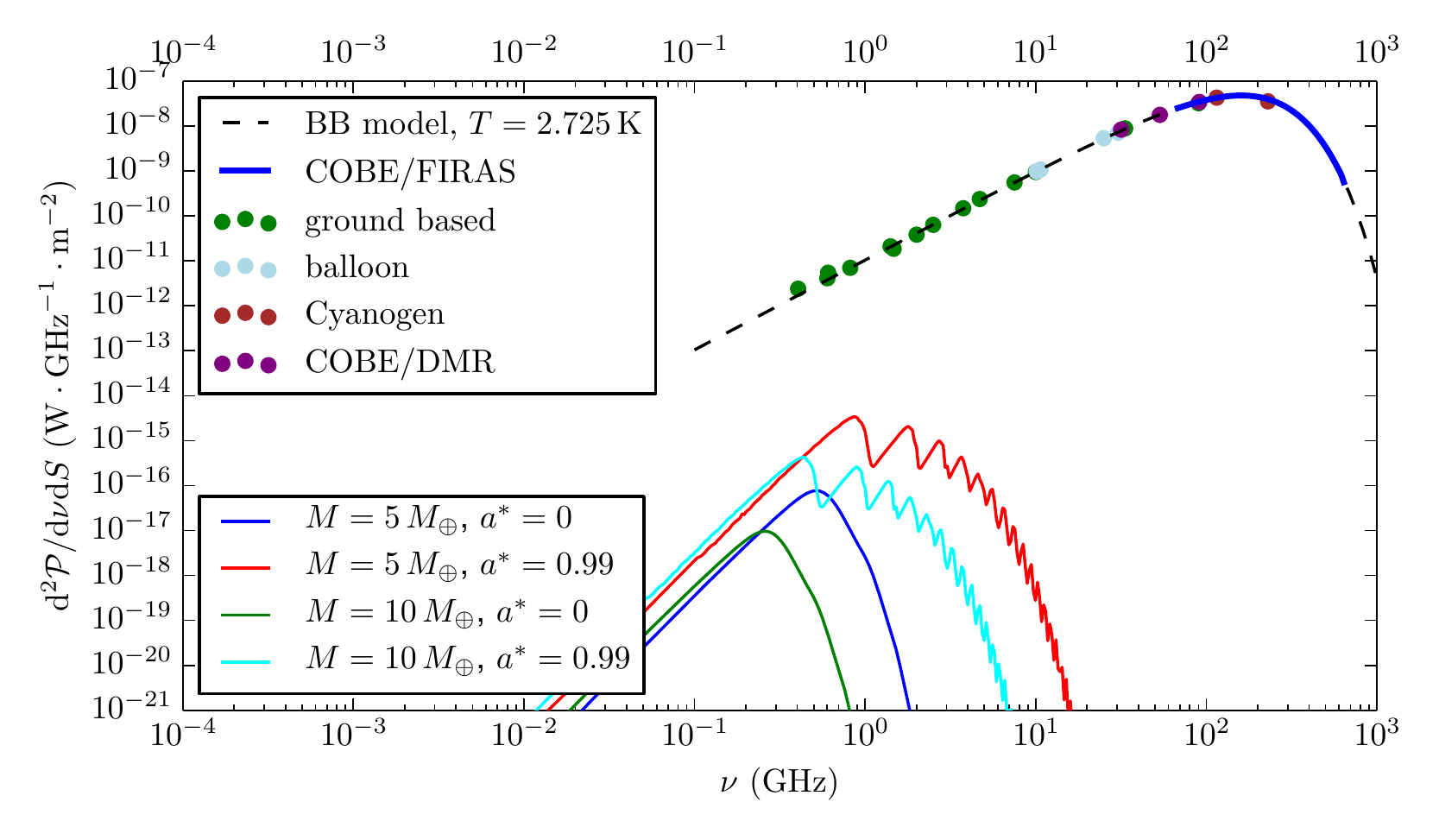}
    \caption{Comparison between the Planet 9 radio signal and the CMB background. Adapted from \cite{Arbey2020_P9}.}
    \label{fig:P9}
\end{figure}

Fourth, \texttt{BlackHawk} can predict the signals emitted by astrophysical structures in terms of Hawking radiation in various energy bands. These constraints are independent from and complementary to the cosmological and galactic ones. For example, PBHs evaporating in nearby dwarf galaxies such as Leo T would deposit energy -- heat and ionization -- in such structures. Constraints can be set on the abundance of PBHs in these strongly DM-dominated objects \cite{Kim2020,Laha2020_2,Coogan2020}. More exotic is the prediction of the radio ($\gamma$) and gravitational waves ($g$) power emission of a putative Planet 9, would it be a PBH of $M_{\rm P9}\sim 10\,M_\oplus$ \cite{Arbey2020_P9}
\begin{equation}
    \frac{\d^2\mathcal{P}_{\gamma/g}}{\d \nu \d S} = \frac{1}{4\pi r\mrm{S}^2}E\frac{\d^2 N_{\gamma/g}}{\d t\d \nu}\,,
\end{equation}
where $r\mrm{S}$ is the Planet 9 Schwarzschild radius. Prediction of the signal in terms of radio photons is given in Fig.~\ref{fig:P9}.

Interestingly, most of these studies used the fact that \texttt{BlackHawk} allows the user to compute the spectra for realistic spinning BHs and extended mass functions. This is the main improvement of this public code over previous constraints, beside the more precise computation of the Hawking radiation rates.

\section{Conclusion}
\label{sec:prospects}

We have summarized here the context of development of the public code \texttt{BlackHawk}, which provides a powerful and versatile tool to compute the Hawking radiation spectra of any distribution of BHs. We have described the basics of Hawking radiation and the main capabilities of \texttt{BlackHawk} and given an overview of the studies this code has already helped to pursue. \texttt{BlackHawk} is not a fixed entity and receives regularly updates to correct bugs and improve its smoothness. Moreover, work is invested towards a major update that will enhance its capabilities: new BH solutions as well as a better low-energy treatment and additional particles will be added in the near future.

\acknowledgments

We would like to thank the organizers of the TOOLS 2020 conference for giving us the opportunity to present our work.

\bibliographystyle{JHEP}
\bibliography{biblio}

\end{document}